\documentclass[a4paper,oneside]{amsart}

\RequirePackage{amsmath}
\RequirePackage{bm}
\RequirePackage{amssymb}
\RequirePackage{upref}
\RequirePackage{amsthm}
\RequirePackage{enumerate}
\RequirePackage{pb-diagram}
\RequirePackage{amsfonts}
\RequirePackage[mathscr]{eucal}
\RequirePackage{verbatim}
\RequirePackage{xr}
\RequirePackage{graphicx}
\RequirePackage[pdftex]{floatflt}
\usepackage{calc}
\usepackage{xspace}

\newcommand{\cf}{cf.\@\xspace}
\newcommand{\resp}{resp.\@\xspace}

\newcommand{\al}{\alpha}
\newcommand{\bet}{\beta}
\newcommand{\ga}{\gamma}
\newcommand{\de}{\delta }
\newcommand{\e}{\epsilon}

\newcommand{\f}{\varphi}

\newcommand{\ka}{\kappa}
\newcommand{\lam}{\lambda}
\newcommand{\m}{\mu}

\newcommand{\om}{\omega}

\newcommand{\vt}{\vartheta}

\newcommand{\s}{\sigma}

\newcommand{\C}{\varGamma}

\newcommand{\Lam}{\varLambda}

\newcommand{\so}{{\mc S_0}}

\newcommand{\const}{\tup{const}}
\newcommand{\ndash}{\nobreakdash--}

\newcommand{\msp[1]}[1]{\mspace{#1mu}}

\newcommand{\R}[1][n+1]{{\protect\mathbb R}^{#1}}

\newcommand{\N}{{\protect\mathbb N}}

\newcommand{\Z}{{\protect\mathbb Z}}
\newcommand{\eR}{\stackrel{\lower1ex \hbox{\rule{6.5pt}{0.5pt}}}{\msp[3]\R[]}}
\newcommand{\eN}{\stackrel{\lower1ex \hbox{\rule{6.5pt}{0.5pt}}}{\msp[1]\N}}
\newcommand{\eO}{\stackrel{\lower1ex
\hbox{\rule{6pt}{0.5pt}}}{\msc O}}

\newcommand\im{\implies}
\newcommand\ra{\rightarrow}

\newcommand{\un}{\infty}
\newcommand{\A}{\forall}

\newcommand{\uu}{\cup}
\newcommand{\ii}{\cap}
\newcommand{\uuu}{\bigcup}

\newcommand{\uud}{ \stackrel{\lower 1ex \hbox {.}}{\uu}}
\newcommand{\uuud}[1]{ \stackrel{\lower 1ex \hbox {.}}{\uuu_{#1}}}
\newcommand\su{\subset}
\newcommand\Su{\Subset}

\newcommand{\sminus}[1][28]{\raise 0.#1ex\hbox{$\scriptstyle\setminus$}}

\newcommand\inn[1]{{\stackrel{\msp[9]\circ}{#1}}}

\newcommand{\wed}{\wedge}

\newcommand\ti{\times }

\newcommand{\abs}[1]{\lvert#1\rvert}

\newcommand{\spd}[2]{\protect\langle #1,#2\protect\rangle}

\newcommand{\tit}{\textit}

\newcommand{\tup}{\textup}

\newcommand{\mc}{\protect\mathcal}
\newcommand{\msc}{\protect\mathscr}

\providecommand{\bysame}{\makebox[3em]{\hrulefill}\thinspace}

\newcommand{\cq}[1]{\glqq{#1}\grqq\,}

\newcommand{\bt}{\begin{thm}}
\newcommand{\bl}{\begin{lem}}
\newcommand{\bc}{\begin{cor}}
\newcommand{\bd}{\begin{definition}}
\newcommand{\bpp}{\begin{prop}}
\newcommand{\br}{\begin{rem}}
\newcommand{\bn}{\begin{note}}
\newcommand{\be}{\begin{ex}}
\newcommand{\bes}{\begin{exs}}
\newcommand{\bb}{\begin{example}}
\newcommand{\bbs}{\begin{examples}}
\newcommand{\ba}{\begin{axiom}}

\newcommand{\et}{\end{thm}}
\newcommand{\el}{\end{lem}}
\newcommand{\ec}{\end{cor}}
\newcommand{\ed}{\end{definition}}
\newcommand{\epp}{\end{prop}}
\newcommand{\er}{\end{rem}}
\newcommand{\en}{\end{note}}
\newcommand{\ee}{\end{ex}}
\newcommand{\ees}{\end{exs}}
\newcommand{\eb}{\end{example}}
\newcommand{\ebs}{\end{examples}}
\newcommand{\ea}{\end{axiom}}

\newcommand{\bp}{\begin{proof}}
\newcommand{\ep}{\end{proof}}
\newcommand{\eps}{\renewcommand{\qed}{}\end{proof}}

\newcommand{\bal}{\begin{align}}

\newcommand{\bi}[1][1.]{\begin{enumerate}[\upshape #1]}
\newcommand{\bia}[1][(1)]{\begin{enumerate}[\upshape #1]}
\newcommand{\bin}[1][1]{\begin{enumerate}[\upshape\bfseries #1]}
\newcommand{\bir}[1][(i)]{\begin{enumerate}[\upshape #1]}
\newcommand{\bic}[1][(i)]{\begin{enumerate}[\upshape\hspace{2\cma}#1]}
\newcommand{\bis}[2][1.]{\begin{enumerate}[\upshape\hspace{#2\parindent}#1]}
\newcommand{\ei}{\end{enumerate}}

\newcommand\ndots{\raise 0.47ex \hbox {,}\hskip0.06em\cdots %
     \raise 0.47ex \hbox {,}\hskip0.06em} 

\newcommand{\q}{\quad}
\newcommand{\qq}{\qquad}

\newcommand{\hp}{\hphantom}

\newcommand\nd{\noindent}

\newskip\Csmallskipamount                                                
\Csmallskipamount=\smallskipamount
\newskip\Cmedskipamount
\Cmedskipamount=\medskipamount
\newskip\Cbigskipamount
\Cbigskipamount=\bigskipamount

\newcommand\cvs{\vspace\Csmallskipamount}   
\newcommand\cvm{\vspace\Cmedskipamount}

\newskip\csa
\csa=\smallskipamount

\newskip\cma
\cma=\medskipamount

\newskip\cba
\cba=\bigskipamount

\newdimen\spt
\newcommand\citem{\cvs\advance\itemno by
1{(\romannumeral\the\itemno})\hskip3pt}
\newcommand{\bitem}{\cvm\nd\advance\itemno by
1{\bf\the\itemno}\hspace{\cma}}

\newcount\itemno
\newcommand{\las}[1]{\label{S:#1}}

\newcommand{\lae}[1]{\label{E:#1}}
\newcommand{\lat}[1]{\label{T:#1}}
\newcommand{\lal}[1]{\label{L:#1}}
\newcommand{\lad}[1]{\label{D:#1}}

\newcommand{\lar}[1]{\label{R:#1}}

\newcommand{\rs}[1]{Section~\ref{S:#1}}

\newcommand{\rt}[1]{Theorem~\ref{T:#1}}
\newcommand{\rl}[1]{Lemma~\ref{L:#1}}
\newcommand{\rd}[1]{Definition~\ref{D:#1}}

\newcommand{\re}[1]{\eqref{E:#1}}

\newskip\thmskip
\thmskip=\parindent

\newskip\hsk
\newenvironment{hinw}{\labelsep=0pt\begin{list}{}{\labelsep=0pt\itemindent=0pt\labelwidth=0pt\leftmargin=\parindent\rightmargin=0pt\partopsep=\cma}%
\item\it\nopagebreak\nopagebreak}%
{\end{list}}

\newcommand\bh{\begin{hinw}}
\newcommand{\eh}{\end{hinw}}

\newtheoremstyle{normal}
  {\cba}
  {\cba}
  {}
  {\thmskip}
  {\bfseries}
  {.}
  {\hsk}
  {}

\newtheoremstyle{abschnitt}
  {\cba}
  {\cba}
  {}
  {\thmskip}
  {\bfseries}
  {.}
  {\hsk}
  {}

\newtheoremstyle{italic}
  {\cba}
  {\cba}
  {\itshape}
  {\thmskip}
  {\bfseries}
  {.}
  {\hsk}
  {}

\newtheoremstyle{aufgaben}
  {\cba}
  {\cba}
  {}
  {}
  {\normalsize\bfseries}
  {.}
  {\hsk}
  {}

\newtheoremstyle{break}
  {\cba}
  {\cba}
  {\itshape}
  {}
  {\bfseries}
  {.}
  {\newline}
  {}

\swapnumbers
\theoremstyle{italic}
\newtheorem{thm}[subsection]{Theorem}
\newtheorem{lem}[subsection]{Lemma}
\newtheorem{prop}[subsection]{Proposition}
\newtheorem{cor}[subsection]{Corollary}

\theoremstyle{normal}
\newtheorem{rem}[subsection]{Remark}
\newtheorem{definition}[subsection]{Definition}
\newtheorem{example}[subsection]{Example}
\newtheorem{examples}[subsection]{Examples}
\newtheorem{ex}[subsection]{Exercise}
\newtheorem{note}[subsection]{}
\newtheorem{axiom}[subsection]{Axiom}

\theoremstyle{aufgaben}
\newtheorem{exs}[subsection]{Exercises}

\swapnumbers

\numberwithin{equation}{section}
\numberwithin{figure}{section}

\newenvironment{textequation}[1][0.8]
{\begin{equation}
\begin{aligned}
\begin{minipage}{#1\linewidth}}
{\end{minipage}
\end{aligned}
\end{equation}
\ignorespacesafterend}

\newcommand{\btext}{\begin{textequation}}
\newcommand{\etext}{\end{textequation}}

\usepackage[german,english]{babel}
\usepackage{graphicx}
\RequirePackage{amsmath}
\RequirePackage{bm}
\RequirePackage{amssymb}
\RequirePackage{upref}
\RequirePackage{amsthm}
\RequirePackage{enumerate}
\RequirePackage{pb-diagram}
\RequirePackage{amsfonts}
\RequirePackage[mathscr]{eucal}
\RequirePackage{verbatim}
\RequirePackage{xr}
\RequirePackage{graphicx}
\RequirePackage[pdftex]{floatflt}
\usepackage{calc}
\usepackage{xspace}

\RequirePackage{upref}
\RequirePackage{amsthm}
\RequirePackage{enumerate}
\usepackage[mathscr]{eucal}

\usepackage{xr-hyper}

\usepackage{calc}

\newlength{\oddsidemarginlength}
\newlength{\topmarginlength}

\newcounter{numberoflines}
\newcounter{tempcc}
\oddsidemargin=\oddsidemarginlength
\evensidemargin=\oddsidemargin
\topmargin=\topmarginlength
\usepackage{hyperref}

\begin{document}

\flushbottom


\title[Transition from big crunch to big bang]{Transition from big crunch to big bang in brane cosmology}

\author{Claus Gerhardt}
\address{Ruprecht-Karls-Universit\"at, Institut f\"ur Angewandte Mathematik,
Im Neuenheimer Feld 294, 69120 Heidelberg, Germany}
\email{gerhardt@math.uni-heidelberg.de}
\urladdr{http://www.math.uni-heidelberg.de/studinfo/gerhardt/}
\thanks{This work has been supported by the Deutsche Forschungsgemeinschaft.}

%
\subjclass[2000]{35J60, 53C21, 53C44, 53C50, 58J05}
\keywords{Lorentzian manifold, mass, cosmological spacetime, general relativity, inverse mean curvature flow, ARW spacetimes, transition from big crunch to big bang, cyclic universe}
\date{\today}
%


\begin{abstract}
We consider  branes $N=I\ti\so$, where $\so$ is an $n$\ndash dimen\-sional space form,
not necessarily compact, in a Schwarzschild-$\tup{AdS}_{(n+2)}$ bulk
$\mc N$. The branes have a big crunch singularity. If a brane is an ARW space, then, under
certain conditions,  there exists a smooth natural transition flow through the singularity to a
reflected brane
$\hat N$, which has a big bang singularity and which can be viewed as a brane in a reflected
Schwarzschild-$\tup{AdS}_{(n+2)}$ bulk
$\hat{\mc N}$. The joint branes
$N\uu \hat N$ can thus be naturally embedded  in $\R[2]\ti
\so$, hence there exists a second possibility of defining a smooth transition from big crunch
to big bang by requiring that $N\uu\hat N$ forms a
$C^\infty$-hypersurface in $\R[2]\times \so$. This last notion of a smooth transition also
applies to branes that are not ARW spaces, allowing a wide range of possible equations of
state.
\end{abstract}

\maketitle

\tableofcontents

\setcounter{section}{-1}
\section{Introduction}

The problem of finding a smooth transition from a spacetime $N$ with a big crunch to a
spacetime $\hat N$ with a big bang singularity has been the focus of some recent works in
general relativity, see e.g., \cite{khoury:cyclic,steinhardt:cyclic} and the references therein.
For abstract spacetimes, i.e., for spacetimes that are not embedded in a bulk space, it is
even a non-trivial question how to define a smooth transition.

In two recent papers \cite{cg:arw,cg:rw} we studied this problem for a special class of
spacetimes, so-called ARW spaces, and used the inverse mean curvature flow to prove that, 
by reflecting the spaces, a smooth transition from big crunch to big bang is
possible.

\cvm
In this paper we look at branes in a Schwarzschild-$\tup{AdS}_{(n+2)}$ bulk $\mc N$,
where the branes are assumed to lie in the black hole region, i.e., the radial coordinate is the
time function. For those branes that are ARW spaces the transition results from \cite{cg:rw}
can be applied to conclude that a smooth transition flow from a brane $N$ to a properly
reflected brane $\hat N$ exists. However, the assumption that the branes are ARW spaces
reduces the number of possible branes drastically, \cf \rt{0.4} and \rt{0.5}. Fortunately, in
the case of embedded spacetimes, it is possible to define a transition through the singularity
without using the inverse mean curvature flow.

\cvm
Let $\mc N$ be a Schwarzschild-$\tup{AdS}_{(n+2)}$ bulk space with a black hole
singularity in $r=0$. We assume that the radial coordinate $r$ is negative, $r<0$.
Then, by switching the light cone and changing $r$ to $-r$ we obtain a reflected
Schwarzschild-$\tup{AdS}_{(n+2)}$ bulk space $\mc {\hat N}$ with a white hole
singularity in $r=0$. These two bulk spacetimes can be pasted together such that
$\mc N\uu \mc{\hat N}$ is a smooth manifold, namely, $\R[2]\ti \so$, which has a
metric singularity in $r=0$. In the black (white) hole region $r$ is the time function
and it is smooth across the singularity.

\cvm
Now, let us consider branes $N$ in the black hole region of $\mc N$. These branes
need not to be ARW spaces, they are only supposed to satisfy the first 
 of the five assumptions imposed for ARW spaces. We call
those branes to be of class $(B)$, \cf \rd{0.1}.

The relation between the geometry of the branes and physics is governed by the
Israel junction condition.

We shall prove existence and transition through the singularity only for single branes, but
this does include a two branes or multiple branes configuration--where then each brane has to
be treated separately. 

Moreover, in the equation of state
\begin{equation}
p=\tfrac\om{n}\rho
\end{equation}
 variable $\om$ are allowed
\begin{equation}
\om=\om_0+\lam,\q\om_0=\const,
\end{equation}
where $\lam=\lam(\log(-r))$ is defined in the bulk. 

\cvm
Branes of class $(B)$ exist in the whole black hole region of $\mc N$, they stretch
from $r=0$, the black hole singularity, to $r=r_0$, the event horizon.

\cvm
The branes whose existence is proved in \rt{3.1} automatically
have a big crunch singularity in $r=0$, since this is the \tit{initial condition} for the
ordinary differential equation that has to be solved. The branes are given by an
embedding
\begin{equation}
y(\tau)=(r(\tau),t(\tau),x^i),\q -a<\tau<0,
\end{equation}
with a big crunch singularity in $\tau=0$ such that $r(0)=0$.

From the modified Friedmann equation we shall deduce that the limit
\begin{equation}
\lim_{\tau\ra 0}t(\tau)=t_0
\end{equation}
exists and without loss of generality we shall suppose $t_0=0$.

\cvm
We then shall define a brane $\hat N\su \mc {\hat N}$ by reflection
\begin{equation}
\hat y(\tau)=(-r(-\tau),-t(-\tau),x^i),\q 0<\tau<a.
\end{equation}
The two branes $N,\hat N$ can be pasted together to yield at least a Lipschitz
hypersurface in $\R[2]\ti\so$. If this hypersurface is of class $C^\un$, then we shall
speak of a smooth transition from big crunch to big bang.

\cvm
To  prove the smoothness we reparametrize $N\uu\hat N$ by using $r$ as a new
parameter instead of $\tau$. The old brane $N$ is then expressed as
\begin{equation}
y(r)=(r,t(r),x^i),\q r<0,
\end{equation}
and the reflected $\hat N$ as
\begin{equation}
\hat y(r)=(r,-t(-r),x^i),\q r>0.
\end{equation}

Hence, $N\uu\hat N$ is a smooth hypersurface, if $\tfrac{dt}{dr}$ is a smooth and
even function in $r$ in a small neighbourhood of $r=0$, $-\e<r<\e$. 

If $\tfrac{dt}{dr}$ is not an even function in $r$, then $N\uu\hat N$ will not be
$C^\un$ hypersurface. Examples can easily be constructed.

\cvm
For branes, that are ARW spaces, the transition flow provides an alternate criterion for a
smooth transition through the singularity. The reflected brane $\hat N$ that is used in this
process is the same as in the above description of pasting together the two embeddings. 

A
version of the modified Friedmann equation, equation \re{1.18}, plays a central role in
determining if the transition is smooth, namely, the quantity $f'e^{\tilde\ga f}$ has to be a
smooth and even function in the variable $e^{\tilde \ga f}$ in order that the transition flow
is smooth, and a smooth and even function in the variable $r=-e^f$, if the joint embedding
is to represent a smooth hypersurface in $\R[2]\ti \so$.

\cvm
The metric in the bulk space $\mc N$ is given by by
\begin{equation}
d\tilde s^2=-\tilde h^{-1} dr^2 +\tilde hdt^2 + r^2\s_{ij}dx^idx^j,
\end{equation}
where $(\s_{ij})$ is the metric of an $n$- dimensional space form $\so$, the radial
coordinate
$r$ is assumed to be negative, $r<0$, and $\tilde h(r)$ is defined by
\begin{equation}
\tilde h=m(-r)^{-(n-1)}+\tfrac2{n(n+1)}\Lam r^2 -\tilde\ka,
\end{equation}
where $m>0$ and $\Lam\le 0$ are constants, and $\tilde\ka=-1,0,1$ is the
curvature of $\so$. We note that we assume that there is a black hole region, i.e., if
$\Lam=0$, then we have to suppose $\tilde\ka =1$.

\cvm
We consider branes $N$ contained in the black hole region $\{r_0<r<0\}$. $N$ will
be a globally hyperbolic spacetime $N=I\ti \so$ with metric
\begin{equation}\lae{0.3}
d\bar s^2=e^{2f}(-(dx^0)^2+\s_{ij}dx^idx^j)
\end{equation}
such that
\begin{equation}
f=f(x^0)=\log(-r(x^0)).
\end{equation}

We may assume that the time variable $x^0=\tau$ maps $N$ on the interval
$(-a,0)$. In $\tau=0$ we have a big crunch singularity induced by the black hole.

\cvm
The relation between geometry and physics is governed by the Israel junction
conditions
\begin{equation}\lae{0.5}
h_{\al\bet}-H\bar g_{\al\bet}=\ka (T_{\al\bet} -\s\bar g_{\al\bet}),
\end{equation}
where $h_{\al\bet}$ is the second fundamental form of $N$,
$H=\bar g^{\al\bet}h_{\al\bet}$ the mean curvature,
$\ka\ne 0$ a constant,
$T_{\al\bet}$ the stress energy tensor of a perfect fluid with an equation of state
\begin{equation}
p=\tfrac\om{n}\rho,
\end{equation}
 and $\s$ the tension of the brane.

One of the parameters used in the definition of  $(n+1)$-dimensional ARW spaces is a
positive constant $\tilde \ga$, which is best  expressed as
\begin{equation}\lae{0.7}
\tilde\ga =\tfrac12 (n+\tilde\om -2).
\end{equation}
If $N$ would satisfy the Einstein equation of a perfect fluid with an equation of state
\begin{equation}\lae{0.8}
p=\tfrac{\tilde\om}n\rho,
\end{equation}
then $\tilde\ga$ would be defined with the help of $\tilde\om$ in \re{0.8}, \cf
\cite[Section 9]{cg:arw}.

ARW spaces with compact $\so$ also have a future mass $\tilde m>0$, which is
defined by
\begin{equation}\lae{0.9}
\tilde m=\lim\int_MG_{\al\bet}\nu^\al\nu^\bet e^{\tilde\om f},
\end{equation}
where $G_{\al\bet}$ is the Einstein tensor and where the limit is defined with respect
to a sequence of spacelike hypersurfaces running into the future singularity, \cf
\cite{cg:mass}. 

For ARW spaces with non-compact $\so$ we simply call the limit
\begin{equation}
\lim\abs{f'}^2e^{\tilde\ga f}=\tilde m,
\end{equation}
which exists by definition and is positive, mass.

\cvm
The most general branes that we consider are branes of class $(B)$, they are supposed to
satisfy only the first of the five conditions that are imposed on  ARW spaces,
\cf \rd{0.3}. 

Let us formulate this condition as

\bd\lad{0.1}
A globally hyperbolic spacetime $N$, $N=I\times \so$, $I=(a,b)$, the metric of which
satisfies \re{0.3}, with $f=f(x^0)$, is said to be of class $(B)$, if there exist positive
constants $\tilde\ga$ and $\tilde m$ such that
\begin{equation}
\lim_{\tau\ra b}\abs{f'}^2e^{2\tilde\ga f}=\tilde m>0.
\end{equation}
We also say that $N$ is of class $(B)$ with constant $\tilde\ga$ and call $\tilde m$
the mass of $N$, though, even in the case of compact $\so$, the relation \re{0.9} is
defined  only under special circumstances. 
\ed

The time function in spacetimes of class $(B)$ has finite future range, \cf
\cite[Lemma 3.1]{cg:arw}, thus  we may---and shall---assume that $b=0$ and
$I=(-a,0)$.

\cvm
By considering branes of class $(B)$ instead of ARW spaces a larger range of
equation of states is possible
\begin{equation}
p=\tfrac\om{n}\rho.
\end{equation}
We shall also consider variable $\om$.

\bl\lal{0.2}
Let $T_{\al\bet}$ be the divergence free stress energy tensor of a perfect fluid in
$N$ with an equation of state
\begin{equation}
p=\tfrac\om{n}\rho,
\end{equation}
where $\om=\om_0+\lam(f)$ and $\om_0=\const$. Assume that $\lam$ is
smooth satisfying
\begin{equation}\lae{0.14}
\lim_{t\ra-\un}\lam(t)=0
\end{equation}
and let $\tilde\mu=\tilde\mu(f)$ be a primitive of $\lam$ such that
\begin{equation}\lae{0.15}
\lim_{t\ra -\un}\tilde\mu(t)=0.
\end{equation}
Then $\rho$ satisfies the conservation law
\begin{equation}\lae{0.16}
\rho=\rho_0e^{-(n+\om_0)f-\tilde\mu},
\end{equation}
where $\rho_0$ is a constant.
\el

A proof will be given in \rs{1}.

\cvm
\br\lar{0.3}
Since the branes, we shall consider, always satisfy the assumptions
\begin{equation}
\lim_{\tau\ra 0}f=-\un,\q -f'>0,
\end{equation}
it is possible to define $\mu=\mu(r)$, $r=-e^f$, by
\begin{equation}\lae{0.18}
\mu(r)=\tilde\mu(\log(-r)).
\end{equation}
We also call $\mu$ a primitive of $\lam$.
\er

\cvm
The main results of this paper can now be summarized in the following four theorems.

\bt\lat{0.4}
Let $N$ be a brane contained in the black hole region of $\mc N$. Let $n\ge
3$ and assume that $\lam=\lam(t)$ satisfies
\begin{equation}\lae{a0.26}
\abs{D^m\lam(t)}\le c_m\qq\A\; m\in\N.
\end{equation}

\cvm
\tup{(i)} If $\s\ge 0$, then $\tilde\ga=\tfrac12(n-1)$ is the only possible value such
that $N$ is an \tup{ARW} space.

\cvm
\tup{(ii)} On the other hand, if we set $\tilde\ga =\tfrac12(n-1)$, 
 then $N$ is
an
\tup{ARW} space if and only if the following conditions are satisfied
\begin{equation}\lae{0.11}
\om_0=-\tfrac{n-1}2\q \tup{and}\q \begin{cases}
\s=0,& \tup{if}\q 3< n\in\N,\\
\s\in\R[],& \tup{if}\q n=3,
\end{cases}
\end{equation}
\begin{equation}\lae{a0.28}
\abs{D^m_t\lam}\le c_m e^{(n-1)t}\qq\A\; m\in\N
\end{equation}
and
\begin{equation}\lae{a0.29}
\lim_{t\ra -\un}\lam(t) e^{-(n-1)t}
\end{equation}
exists,
or
\begin{equation}
\om_0\le -(n-1)\q\tup{and}\q \s\in\R[],\;  \A\; 3\le n\in\N.
\end{equation}

If the condition \re{0.11} holds, then the mass $\tilde m$ of $N$ is larger than $m$
\begin{equation}
\tilde m=m +\tfrac{\ka^2}{n^2}\rho_0^2,
\end{equation}
where $\rho_0$ is an integration constant of $\rho e^{(n+\om_0)f}e^{\tilde\mu}$. In the
other cases we have $\tilde m=m$.

\cvm
\tup{(iii)} There exists a smooth transition flow from $N$ to a reflected brane $\hat N$, if
the primitive $\mu$ can be viewed as a smooth and even function in $e^{\tilde\ga f}$, and
provided the following conditions are valid
\begin{equation}
n=3,\q \om_0=-\tfrac{n-1}2,\q \s\in\R[],
\end{equation}
or
\begin{equation}
n>3,\q \om_0=-\tfrac{n-1}2,\q \Lam=\s=0,\q \tilde\ka =1,
\end{equation}
or
\begin{equation}
n=3,\q \om_0=-m(n-1)+1,\;  2\le m\in\N,\q \s\in\R[],
\end{equation}
or
\begin{equation}
n>3,\q \om_0=-m(n-1)+1,\; 2\le m\in\N,\q \tfrac2{n(n+1)}\Lam=-\s^2,
\end{equation}
where in case $\Lam=\s=0$, we have to assume $\tilde\ka=1$.

\cvm
\tup{(iv)}  Since an ARW brane is also a brane of class $(B)$, the smooth transition results
from \rt{0.7}, \tup{(i)}, are valid, if the primitive $\mu$ can be viewed as a smooth and
even function in the variable $r=-e^f$.

\cvm
\tup{(v)} For the specified values of $\om_0$ and $\s$ the branes do actually exist.
\et

\bt\lat{0.5}
Assume that $\lam$ vanishes in a neighbourhood of $-\un$. Then a brane $N\su
\hat{\mc N}$ is an
\tup{ARW} space with
$\tilde\ga\ne\tfrac12(n-1)$ if and only if $\tilde\ga =n$, $\om_0 =1$, and
\begin{equation}
\s=-\tfrac m{2\rho_0}.
\end{equation}
The mass $\tilde m$ of $N$ is then equal to
\begin{equation}
\tilde m=\tfrac{\ka^2}{n^2}\rho_0^2.
\end{equation}

\tup{(i)} There exists a smooth transition flow, if 
\begin{equation}
\tfrac2{n(n+1)}\Lam =-\s^2.
\end{equation}

\tup{(ii)} Since $N$ is also a brane of class $(B)$, the smooth transition result from
\rt{0.7},
\tup{(ii)}, is valid, i.e., the joint branes $N\uu\hat N$ form a $C^\infty$-hypersurface in
$\R[2]\ti \so$, if $n\ge 3$ is odd, and if $\mu$ can be viewed as a smooth and even
function  in $r$.

A brane with the specified values does actually exist.
\et

\bt
A brane $N\su\mc N$ satisfying an equation of state with $\om=\om_0+\lam$,
where $\lam$ satisfies the conditions of \rl{0.2},   is of class
$(B)$ with constant
$\tilde\ga>0$ if and only if
\begin{equation}
\tilde\ga =\tfrac{n-1}2\q\tup{and}\q \om_0\le -\tfrac{n-1}2,
\end{equation}
or
\begin{equation}
\tilde\ga=n+\om_0-1\q\tup{and}\q \om_0>-\tfrac{n-1}2.
\end{equation}
In both cases the tension $\s\in\R[]$ can be arbitrary.
Branes with the specified values do actually exist.
\et

\bt\lat{0.7}
Let $N$ be a brane of class $(B)$ as described in the preceding theorem and assume that
the corresponding function $\mu$ satisfies the conditions stated in \re{2.10} \resp
\re{2.11}, which more or less is tantamount to requiring that $\mu$ is smooth and even as
a function of $r$. Then
$N$ can be reflected to yield a brane $\hat N\su \mc{\hat N}$. The joint branes $N\uu
\hat N$ form a $C^\un$- hypersurface in $\R[2]\ti\so$ provided the following
conditions are valid

\cvm
\tup{(i)} If $\tilde\ga=\tfrac{n-1}2$ and $\om_0\le -\tfrac{n-1}2$, then the
relations
\begin{equation}
\tfrac{n-1}2\; odd,\q \om_0\; odd,\q \tup{and}\q \s\in\R[],
\end{equation}
or
\begin{equation}
n\; odd,\q \om_0\in\Z,\q \tup{and}\q \s=0
\end{equation}
should hold.

\cvm
\tup{(ii)} If $\tilde\ga=n+\om_0-1$ and $\om_0>-\tfrac{n-1}2$, then the relations
\begin{equation}
n\; odd,\q \om_0\; odd,\q \tup{and}\q \s\in\R[],
\end{equation}
or
\begin{equation}
n\; odd,\q \om_0\in\Z,\q \tup{and}\q \s=0
\end{equation}
should be valid.
\et

\cvm
For the convenience of the reader we repeat the definiton of ARW spaces, slightly
modified to include the case of non-compact $\so$.
\bd\lad{0.3}
A globally hyperbolic spacetime $N$, $\dim N=n+1$, is said to be \tit{asymptotically
Robertson-Walker} (ARW) with respect to the future, if a future end of $N$, $N_+$,
can be written as a product $N_+=[a,b)\times \so$, where $\so$ is a 
Riemannian space, and there exists a future directed time function $\tau=x^0$ such
that the metric in $N_+$ can be written as
\begin{equation}\lae{0.16b}
d\breve s^2=e^{2\tilde\psi}\{-{(dx^0})^2+\s_{ij}(x^0,x)dx^idx^j\},
\end{equation}
where  $\so$ corresponds to $x^0=a$, $\tilde\psi$ is of the form
\begin{equation}
\tilde\psi(x^0,x)=f(x^0)+\psi(x^0,x),
\end{equation}
and we assume that there exists a positive constant $c_0$ and a smooth
Riemannian metric $\bar\s_{ij}$ on $\so$ such that
\begin{equation}
\lim_{\tau\ra b}e^\psi=c_0\q\wed\q \lim_{\tau\ra b}\s_{ij}(\tau,x)=\bar\s_{ij}(x),
\end{equation}
and
\begin{equation}
\lim_{\tau\ra b}f(\tau)=-\un.
\end{equation}

\cvm
Without loss of generality we shall assume $c_0=1$. Then $N$ is ARW with
respect to the future, if the metric is close to the Robertson-Walker metric
\begin{equation}\lae{0.20}
d\bar s^2=e^{2f}\{-{dx^0}^2+\bar\s_{ij}(x)dx^idx^j\}
\end{equation}
near the singularity $\tau =b$. By \tit{close} we mean that the derivatives of arbitrary order with respect to space and time of the
conformal metric $e^{-2f}\breve g_{\al\bet}$ in \re{0.16b} should converge  to the
corresponding derivatives of the conformal limit metric in \re{0.20}, when $x^0$
tends to $b$. We emphasize that in our terminology Robertson-Walker metric does not
 necessarily imply that
$(\bar\s_{ij})$ is a metric of constant curvature, it is only the spatial metric of a
warped product.

\cvm
We assume, furthermore, that $f$ satisfies the following five conditions
\begin{equation}
-f'>0,
\end{equation}
there exists $\tilde\om\in\R[]$ such that
\begin{equation}\lae{0.22}
n+\tilde\om-2>0\q\wed\q \lim_{\tau\ra b}\abs{f'}^2e^{(n+\tilde\om-2)f}=\tilde
m>0.
\end{equation}
Set $\tilde\ga =\frac12(n+\tilde\om-2)$, then there exists the limit
\begin{equation}\lae{0.23}
\lim_{\tau\ra b}(f''+\tilde\ga \abs{f'}^2)
\end{equation}
and
\begin{equation}\lae{0.24}
\abs{D^m_\tau(f''+\tilde\ga \abs{f'}^2)}\le c_m \abs{f'}^m\qq
\A\, m\ge 1,
\end{equation}
as well as
\begin{equation}\lae{0.25}
\abs{D_\tau^mf}\le c_m \abs{f'}^m\qq\A\, m\ge 1.
\end{equation}

\cvm
If $\so$ is compact, then we call $N$ a \tit{normalized} ARW spacetime, if
\begin{equation}
\int_{\so}\sqrt{\det{\bar\s_{ij}}}=\abs{S^n}.
\end{equation}
\ed

\br
The special branes we consider are always Robertson-Walker spaces, i.e., in order to
prove that they are also ARW spaces we only have to show that $f$ satisfies the
five conditions stated above.
\er

\section{The modified Friedmann equation}\las{1}

The Israel junction condition \re{0.5} is equivalent to
\begin{equation}
h_{\al\bet}=\ka (T_{\al\bet}-\tfrac1n T\bar
g_{\al\bet}+\tfrac\s{n}\bar g_{\al\bet}),
\end{equation}
where $T=T^\al_\al$.

Assuming the stress energy tensor to be that of a perfect fluid
\begin{equation}
T^0_0=-\rho,\q T^\al_i=p\de^\al_i,
\end{equation}
satisfying an equation of state
\begin{equation}
p=\tfrac\om{n}\rho,
\end{equation}
we finally obtain
\begin{equation}\lae{1.4}
h_{ij}=\tfrac\ka{n}(\rho+\s)\bar g_{ij}
\end{equation}
for the spatial components of the second fundamental form.

\cvm
Let us label the coordinates in $\mc N$ as $(y^a)=(y^r,y^t,y^i)\equiv (r,t,x^i)$.
Then we consider embeddings of the form
\begin{equation}
y=y(\tau,x^i)=(r(\tau),t(\tau),x^i)
\end{equation}
$x^0=\tau$ should be the time function on the brane which is chosen such that the
induced metric can be written as
\begin{equation}
d\bar s^2=r^2(-(dx^0)^2+\s_{ij}dx^idx^j).
\end{equation}

We also assume that $\dot r>0$. Notice that  $r<0$, so that $(x^\al)$ is a
future oriented coordinate system on the brane. If we set $f=\log (-r)$, then the
induced metric has the form as indicated in \re{0.3}. 

Let us point out that this choice of $\tau$ implies the relation
\begin{equation}\lae{1.7}
r^2=\tilde h^{-1}\dot r^2 -\tilde h\abs{t'}^2,
\end{equation}
since
\begin{equation}
\bar g_{00}= \spd{\dot y}{\dot y},
\end{equation}
where we use a dot or a prime to indicate differentiation with respect to $\tau$ unless
otherwise specified.

Since the time function in ARW spaces or in spaces of class $(B)$ has a finite future
range,
\cf
\cite[Lemma 3.1]{cg:arw}, we assume without loss of generality that the embedding
is defined in
$I\times \so$ with $I=(-a,0)$.

The only non-trivial tangent vector of $N$ is
\begin{equation}
 y'=( r', t', 0\ldots,0),
\end{equation}
and hence a covariant normal $(\nu_a)$ of $N$ is given by
\begin{equation}
\lam (-t',r',0,\ldots,0),
\end{equation}
where $\lam$ is a normalization factor, and the contravariant normal vector is given
by
\begin{equation}\lae{1.11}
(\nu^a)=-r^{-1}( \tilde h\tfrac{dt}{d\tau},\tilde h^{-1}\dot r, 0,\ldots, 0),
\end{equation}
in view of \re{1.7}.

The normal vector $\nu$ of the brane is spacelike, i.e., the Gau{\ss} formula reads
\begin{equation}\lae{1.12}
y_{\al\bet}=-h_{\al\bet}\nu,
\end{equation}
 we refer to \cite[Section 2]{cg:indiana} for our conventions. 

We also emphasize
that we have neither specified the sign nor the actual value of $\ka$, i.e., it is
irrelevant which normal we use in the Gau{\ss} formula.

\cvm
To determine $h_{ij}$ we use
\begin{equation}\lae{1.13}
\begin{aligned}
y_{ij}^t&=y_{,ij}^t+\tilde\C_{bc}^ty^b_iy^c_j-\bar\C^\ga_{ij}y^t_\ga=-r^{-1}\dot
r \tfrac{dt}{d\tau}\s_{ij},
\end{aligned}
\end{equation}
and we conclude
\begin{equation}
h_{ij}=-\tilde h\tfrac{dt}{d\tau}\s_{ij},
\end{equation}
in view of \re{1.12}, \re{1.13} and the assumption $\dot r\ne 0$, and from \re{1.4}
we further deduce
\begin{equation}\lae{1.15}
-\tilde h r^{-2}\tfrac{dt}{d\tau}=\tfrac\ka{n}(\rho+\s),
\end{equation}
or, by taking \re{1.7} into account,
\begin{equation}\lae{1.16}
\abs{f'}^2-\tilde h=\tfrac{\ka^2}{n^2}(\rho+\s)^2e^{2f}.
\end{equation}

This is the modified Friedmann equation.

\bh
Branes that are ARW spaces
\eh

Let us first consider branes that are ARW spaces.

\cvm
Since the bulk is an Einstein space, the left-hand side of \re{0.5} is divergence free, as
can be easily deduced with the help of the Codazzi equation, i.e., $T_{\al\bet}$ is also
divergence free, and hence $\rho$ satisfies the conservation law
\begin{equation}\lae{1.17}
\rho e^{(n+\om_0)f}e^{\tilde\mu}=\const=\rho_0,
\end{equation}
\cf \rl{0.2}; a proof will be given later.

\cvm
In order to find out under which conditions the brane is an ARW space, we distinguish
between the cases $\s\ge 0$ and $\s<0$. The latter choice violates the
approximation of the classical Friedmann equation for small $\rho$.

\bl
Let $\s\ge 0$, then $\tilde\ga=\tfrac12(n-1)$ is the only possible value such that
$N$ can be an \tup{ARW} space.
\el

\bp
From \re{1.16} and \re{1.17} we derive
\begin{equation}\lae{1.18}
\begin{aligned}
\abs{f'}^2e^{2\tilde\ga f}&=m e^{(2\tilde\ga  -(n-1))f}+\tfrac2{n(n+1)}\Lam
e^{2(\tilde\ga+1)f}-\tilde\ka e^{2\tilde\ga f}\\[\cma]
&\hp{=}+\tfrac{\ka^2}{n^2}(\rho_0^2e^{2(\tilde\ga+1
-(n+\om_0))f}e^{-2\tilde\mu}\\[\cma] &\qq
+
2\s\rho_0e^{(2\tilde\ga+2-(n+\om_0))f}e^{-\tilde\mu}+\s^2e^{2(\tilde\ga +1)f}).
\end{aligned}
\end{equation}

Differentiating both sides and dividing by $2f'$ yields
\begin{equation}\lae{1.19}
\begin{aligned}
&\qq(f''+\tilde\ga\abs{f'}^2)e^{2\tilde\ga f}= \\[\cma]
& m(\tilde\ga
-\tfrac{(n-1)}2)e^{(2\tilde\ga -(n-1))f}+\tfrac{2(\tilde\ga +1)}{n(n+1)}\Lam
e^{2(\tilde\ga +1)f}-\tilde\ga\tilde\ka e^{2\tilde\ga f}\\[\cma]
&+\tfrac{\ka^2}{n^2}(\rho_0^2(\tilde\ga+1 -(n+\om_0))e^{2(\tilde\ga
+1-(n+\om_0))f}e^{-2\tilde\mu}\\[\cma]
&\qq-\lam\rho_0^2 e^{2(\tilde\ga
+1-(n+\om_0))f}e^{-2\tilde\mu}\\[\cma]
&\qq+\s\rho_0(2\tilde\ga +2-(n+\om_0))e^{(2\tilde\ga
+2-(n+\om_0))f}e^{-\tilde\mu}\\[\cma]
&\qq-\lam\s\rho_0 e^{(2\tilde\ga
+2-(n+\om_0))f}e^{-\tilde\mu}\\[\cma]
&\qq+\s^2(\tilde\ga +1)e^{2(\tilde\ga +1)f}).
\end{aligned}
\end{equation}

If $N$ is an ARW space, then the left-hand side of \re{1.18} has to converge to a
positive constant, if $\tau$ goes to zero, and $f''+\tilde\ga\abs{f'}^2$ has to
converge to a constant.

Thus we  deduce that all exponents of
$e^f$ on the right-hand side of
\re{1.18} have to be  non-negative. Dividing now equation \re{1.19} by $e^{2\tilde\ga f}$,
and using the fact that  the terms involving $\lam$ 
can be neglected, since $\lam$ vanishes sufficiently fast near $-\un$,  we see that the
coefficients of all powers of
$e^f$, which have the common factor
$\tfrac{\ka^2}{n^2}$, are non-negative, hence we must have $\tilde\ga
=\tfrac12(n-1)$, for otherwise we get a contradiction.
\ep

\bl\lal{1.2}
Let $\tilde\ga=\tfrac12(n-1)$, $\s\in\R[]$ and assume that \re{a0.26} is valid. Then $N$ is
an
\tup{ARW} space if and only if the following conditions are satisfied
\begin{equation}\lae{1.20}
\om_0=-\tfrac{n-1}2\q \tup{and}\q \begin{cases}
\s=0,& \tup{if}\q 3< n\in\N,\\
\s\in\R[],& \tup{if}\q n=3,
\end{cases}
\end{equation}
and the relations \re{a0.28} and \re{a0.29} hold,
or
\begin{equation}\lae{1.21}
\om_0\le -(n-1)\q\tup{and}\q \s\in\R[],\;  \A\; 3\le n\in\N.
\end{equation}
\el

\bp
Let $n\ge 3$ and $\s\in\R[]$ be arbitrary. If the left-hand side of \re{1.18}
converges, then the exponents of the terms with the common factor
$\tfrac{\ka^2}{n^2}$ have to be non-negative, since the exponents of the terms
with the factor $\rho_0$ can't be both negative and equal, so that they might cancel
each other. Hence there must hold
\begin{equation}
\om_0\le -\tfrac{n-1}2.
\end{equation}

Moreover, after didividing \re{1.19} by $e^{2\tilde\ga f}$ we see that either $\om_0=
-\tfrac{n-1}2$ or
\begin{equation}
n+\om_0\le 1,
\end{equation}
i.e.,
\begin{equation}
\om_0\le -(n-1).
\end{equation}

In case $\om_0=-\tfrac{n-1}2$, we deduce from \re{1.19} that either $\s=0$ or
\begin{equation}
0\le 2-n+\tfrac{n-1}2=-\tfrac{n-3}2,
\end{equation}
i.e., $n=3$ must be valid, if $\s\ne 0$.

\cvm
If these necessary conditions are satisfied, then we can express
$f''+\tilde\ga\abs{f'}^2$ in the form
\begin{equation}
\begin{aligned}
f''&+\tilde\ga\abs{f'}^2=\tfrac1n\Lam e^{2f}-\tilde\ga\tilde\ka\\[\cma]
&\q+\tfrac{\ka^2}{n^2}(-\lam\rho_0^2 e^{-(n-1)f}
e^{-2\tilde\mu}+2\s\rho_0e^{-\tilde\mu}\\[\cma]
&\qq\q\; \;  -\lam\s\rho_0e^{-\tilde\mu}+\s^2(\tilde\ga +1)e^{2f})
\end{aligned}
\end{equation}
 in case $\om_0=-\tfrac{n-1}2$, where we note that $\s=0$, if
$n>3$, and
$\tilde\ga =1$, if
$n=3$, and in the form
\begin{equation}
\begin{aligned}
f''&+\tilde\ga\abs{f'}^2=\tfrac1n\Lam e^{2f}-\tilde\ga\tilde\ka\\[\cma]
&\q+\tfrac{\ka^2}{n^2}(c_1 e^{2\e_1 f}e^{-2\tilde\mu}-\lam c_2 e^{2\e_1 f}
e^{-2\tilde\mu}+c_3  e^{(\e_1+1)f}e^{-\tilde\mu}\\[\cma]
&\qq\q\; \;  -\lam c_4 e^{(\e_1+1)f} e^{-\tilde\mu}+c_5 e^{2f})
\end{aligned}
\end{equation}
with constants $c_i, \e_1$, such that $\e_1\ge 0$, if $\om_0\le -(n-1)$.

 Thus the remaining conditions for $f$ in the definition of ARW spaces are
automatically satisfied, in view of the conditions \re{a0.26}, \re{a0.28}, and \re{a0.29}.

On the other hand, it is immediately clear that the conditions in the lemma are also
sufficient provided we have a solution of equation \re{1.16} such that
\begin{equation}\lae{1.28}
\lim_{\tau\ra 0}f=-\un \q\tup{and} \q f'<0.\qedhere
\end{equation}
\ep

The existence of such a solution will be shown in \rt{3.1}.

\cvm
Next let us examine the possibility that $N$ is an ARW space with $\tilde\ga\ne
\tfrac{n-1}2$.

\bl\lal{1.4}
Let $\tilde\ga\ne\tfrac{n-1}2$, and suppose that $\lam$ vanishes in a neighbourhood of
$-\un$.  Then
$N$ is an
\tup{ARW} space with constant
$\tilde\ga$ if and only if $\tilde\ga=n$, $\om=1$, and $\s<0$ is fine tuned to
\begin{equation}\lae{1.32}
\s=-\tfrac m{2\rho_0},
\end{equation}
where $\rho_0$ is the integration constant in \re{1.17}.
\el

\bp
Let $N$ be an ARW space with $\tilde\ga\ne \tfrac{n-1}2$, then we conclude from
\re{1.18} that all exponents of $e^f$ had to be non-negative or
\begin{equation}\lae{1.33}
2-(n+\om_0)=-(n-1),
\end{equation}
i.e., $\om_0=1$, and  $\s$ had to be fine tuned as indicated in \re{1.32}.
If all exponents were non-negative, then we would use \re{1.19} to deduce the same
result as in \re{1.33} with the corresponding value for $\s$.

Hence, in any case we must have $\om_0=1$ and $\s$ as in \re{1.32}. Inserting these
values in \re{1.18} we conclude
\begin{equation}
0=2\tilde\ga +2-2(n+1),
\end{equation}
i.e., $\tilde\ga =n$.

\cvm
The conditions in the lemma are therefore necessary. Suppose they are satisfied, then
we deduce
\begin{equation}
\lim\abs{f'}^2e^{2\tilde\ga f}=\tfrac{\ka^2}{n^2}\rho_0^2
\end{equation}
and
\begin{equation}
f''+\tilde\ga\abs{f'}^2=\tfrac2n\Lam e^{2f}-\tilde\ga\tilde\ka
+\tfrac{\ka^2}{n^2}\s^2(n+1)e^{2f},
\end{equation}
if $f$ is close to $-\un$, from which we immediately infer that $N$ is an ARW space.
\ep

For the existence result we refer to \rt{3.1}.

\br\lar{1.5}
(i) If in \rl{1.2} \resp \rl{1.4} $\om_0$ is equal to the isolated values
$\om_0=-\tfrac{n-1}2$ \resp $\om_0=1$, then the future mass of $N$ is different from
the mass $m$ of $\mc N$, namely, in case $\om_0=-\tfrac{n-1}2$ and
$\tilde\ga=\tfrac{n-1}2$ we get
\begin{equation}
\tilde m=m+\tfrac{\ka^2}{n^2}\rho_0^2>m,
\end{equation}
and in case $\om_0=1$ and $\tilde\ga =n$
\begin{equation}
\tilde m=\tfrac{\ka^2}{n^2}\rho_0^2.
\end{equation}

\cvm
(ii) In the  case $\tilde\ga=\tfrac {n-1}2$, the value of
$\tilde\ga$ is equal to the value that one would get assuming the Einstein equations
were valid in $N$, where the stress energy tensor would be that of a perfect fluid
with an equation of state
\begin{equation}
p=\tfrac{\tilde\om}n\rho
\end{equation}
such that $\tilde\om=1$, since then $N$ would be an ARW space satisfying \re{0.7},
\cf \cite[Section 9]{cg:arw}.

Furthermore, the classical Friedmann equation has the form
\begin{equation}\lae{1.40}
\abs{f'}^2=-\tilde\ka +\tfrac \ka{n}\rho e^{2f}=-\tilde\ka+\tfrac \ka{n}\rho_0
e^{(2-(n+\tilde\om))f}.
\end{equation}

Thus, by identifying the leading powers of $e^f$ on the right-hand side of \re{1.16}
and \re{1.40}, we see that in case $\tilde\om=1$ there should hold
\begin{equation}
-(n+1)=-2(n+\om_0),
\end{equation}
i.e.,
\begin{equation}
\om_0=-\tfrac{n-1}2.
\end{equation}

Therefore, if we want to  interpret the modified Friedmann equation in analogy to the
classical Friedmann equation, then we have to choose $\om_0=-\tfrac{n-1}2$, if the
relation \re{0.7} serves as a guiding principle.
\er

Let us conclude these considerations on branes that are ARW spaces  with the
following lemma

\bl
The branes $N$ which are \tup{ARW} spaces satisfy the timelike convergence
condition near the singularity.
\el

\bp
We apply the relation connecting  the Ricci tensors of two conformal metrics. Let 
\begin{equation}
(\nu^\al)=(1,u^i)\q\tup{and}\q v=1-\s_{ij}u^iu^j\equiv 1-\abs{Du}^2> 0,
\end{equation}
then
\begin{equation}
\begin{aligned}
\bar
R_{\al\bet}\nu^\al\nu^\bet&=R_{ij}u^iu^j-(n-1)(f''-\abs{f'}^2)+v^2(-f''-(n-1)\abs{f'}^2)\\[\cma]
&\ge (n-1)\tilde\ka \abs{Du}^2-(n-2)f''+(n-1)\abs{f'}^2\abs{Du}^2,
\end{aligned}
\end{equation}
hence the result.
\ep

\bh
Branes of class $(B)$
\eh

Next we shall examine branes that are of class $(B)$. Let us start with a proof of
\rl{0.2}

\bp[Proof of \rl{0.2}]
The equation $T^\ga_{0;\ga}=0$ implies
\begin{equation}
0=-\dot\rho -(n+\om)f'\rho.
\end{equation}

If $\rho$ would vanish in a point, then it would vanish identically, in view of Gronwall's
lemma. Hence, we may assume that $\rho>0$, so that
\begin{equation}
(\log\rho)'=-(n+\om_0)f'-\lam f',
\end{equation}
from which the conservation law \re{0.9} follows immediately.
\ep

In the following two lemmata we shall assume that $T_{\al\bet}$ satisfies an
equation of state with $\om=\om_0+\lam$, such that $\lam,\tilde\mu$ and $\mu$
satisfy the relations \re{0.14}, \re{0.15} and \re{0.18}.

\bl\lal{1.7}
Let $\tilde\ga=\tfrac{n-1}2$ and $\om=\om_0+\lam$. Then the brane $N$ is of
class $(B)$ with constant $\tilde\ga$ if and only if
\begin{equation}\lae{1.47}
\om_0\le -\tfrac{n-1}2.
\end{equation}
\el

\bp
The left hand-side of  equation \re{1.18} converges to a positive constant if and only if
$\om_0$ satisfies \re{1.47}.
\ep

\bl\lal{1.8}
Let $\tilde\ga\ne\tfrac{n-1}2$, $\rho>0$, and $\om=\om_0+\lam$. Then the brane
$N$ is of class $(B)$ with constant $\tilde\ga$ if and only if
\begin{equation}\lae{1.49}
\tilde\ga=n+\om_0-1
\end{equation}
and
\begin{equation}\lae{1.50}
\om_0>-\tfrac{n-1}2.
\end{equation}
\el

\bp
(i) Let $N$ be of class $(B)$ with constant $\tilde \ga\ne \tfrac{n-1}2$, then
\begin{equation}
n+\om_0>0
\end{equation}
must be valid, for otherwise we get a contradicition. Therefore
\begin{equation}
2(\tilde\ga+1-(n+\om_0))
\end{equation}
is the smallest exponent on the right-hand side of \re{1.18} and has thus to vanish,
i.e.,
\begin{equation}
\tilde\ga=n+\om_0-1.
\end{equation}

The other exponents have to be non-negative, i.e.,
\begin{equation}\lae{1.54}
2\tilde\ga>n-1
\end{equation}
and
\begin{equation}\lae{1.55}
2\tilde\ga\ge n+\om_0\q\tup{or}\q \s=0.
\end{equation}

The inequality \re{1.54} is equivalent to \re{1.50} which in turn implies \re{1.55}.

\cvm
(ii) The conditions \re{1.49} and \re{1.50} are sufficient, since then
\begin{equation}
2\tilde\ga>n-1
\end{equation}
and
\begin{equation}
\tilde m=\tfrac{\ka^2}{n^2}\rho_0^2.
\end{equation}
\ep

Similarly as before  branes of class $(B)$ do actually exist, \cf \rt{3.1}.

\section{Transition from big crunch to big bang}

\bh
Branes that are ARW spaces
\eh

For a brane $N$ that  is an ARW space, the results of \cite{cg:rw} can be applied. Define a
reflected spacetime $\hat N$ by switching the lightcone and changing the time function to
$\hat x^0=-x^0$, then $N$ and $\hat N$ can be pasted together at the singularity
$\{0\}\times \so$ yielding a smooth manifold with a metric singularity which is a big crunch,
when viewed from $N$, and a big bang, when viewed from $\hat N$. Moreover, there exists a
natural transition flow of class
$C^3$ across the singularity which is defined by rescaling an appropriate inverse mean
curvature flow. 

This transition flow is of class $C^\un$, if the quantity $f'e^{\tilde\ga f}$, or
equivalently, $\abs{f'}^2e^{2\tilde\ga f}$ can be viewed as a smooth and even function in
the variable $e^{\tilde\ga f}$, \cf \cite[the remarks before Theorem 2.1]{cg:rw}. 

For the
branes considered in \rt{0.4} and \rt{0.5} $\abs{f'}^2e^{2\tilde\ga f}$ is a smooth and
even function in $e^{\tilde\ga f}$.

\cvm
In the following we shall see that the reflected spacetime can  be viewed as a brane in the
Schwarzschild-$\tup{AdS}_{(n+2)}$ space $\hat{\mc N}$ which is a reflection of
$\mc N$. $\hat{\mc N}$ is obtained by switching the lightcone in $\mc N$ and by
changing the radial coordinate $r$ to $-r$. The singularity in $r=0$ is then a white
hole singularity.

\cvm
Moreover, ARW branes are also branes of class $(B)$, i.e., the transition results for those
branes, which we shall prove  next, also apply.

\bh
Branes of class (B)
\eh

The brane $N$ is given by
\begin{equation}
y(\tau)=(r(\tau),t(\tau),x^i).
\end{equation}
Since the coordinates $(x^i)$ do not change, let us write $y(\tau)=(r(\tau),t(\tau))$.

We shall see in a moment that $t(\tau)$ converges, if $\tau$ tends to zero, hence,
let us set this limit  equal to zero.

Now we define the  reflection $\hat N$ of $N$ by
\begin{equation}
\hat y(\tau)=(-r(-\tau),-t(-\tau))
\end{equation}
for $\tau>0$. The result is a brane in $\hat{\mc N}$ which can be pasted
continuously to $N$.

The embeddings of $N$ \resp $\hat N$ in $\mc N$ \resp $\hat{\mc N}$ are also
embeddings in $\R[2]\ti\so$, endowed with the Riemannian product metric, and now
it makes sense to ask, if the joint hypersurfaces $N\uu\hat N$ form a smooth
hypersurface in $\R[2]\ti\so.$

The smoothness of $N\uu\hat N$ in $\R[2]\ti\so$ we interpret as a smooth
transition from big crunch to big bang.

\cvm
To prove the smoothness we have to parameterize $y=y(\tau)$ with respect to $r$.
This is possible, since
\begin{equation}
\tfrac{dr}{d\tau}=-f'e^f>0.
\end{equation}

Then we have $y(r)=(r,t(r))$ and we need to examine
\begin{equation}
\tfrac{dt}{dr}=\tfrac{dt}{d\tau}\tfrac{d\tau}{dr}=-\tfrac{dt}{d\tau}\tfrac1{f'e^f}.
\end{equation}

Define
\begin{equation}
\begin{aligned}
\vt(r)&=\tilde h(r) e^{(n-1)f}\\
&=m+\tfrac2{n(n+1)}\Lam e^{(n+1)f}-\tilde\ka e^{(n-1)f},
\end{aligned}
\end{equation}
then we obtain from \re{1.15} and \re{1.17}
\begin{equation}\lae{2.6}
\begin{aligned}
\vt\tfrac{dt}{dr}&=\tfrac\ka{n}(\s+\rho_0
e^{-(n+\om_0)f}e^{-\tilde\mu})\tfrac{e^{nf}}{f'}\\[\cma] &=\tfrac\ka{n}(\s
e^{(n+\tilde\ga)f}+\rho_0 e^{(\tilde\ga -\om_0)f}e^{-\tilde\mu})\tfrac1{f' e^{\tilde\ga
f}}.
\end{aligned}
\end{equation}

This immediately shows that in all cases  $\abs{\tfrac{dt}{dr}}$ is uniformly
bounded, hence $\lim_{r\ra 0} t(r)$ exists.

\cvm
The joint embedding $N\uu \hat N$ in $\R[2]\ti\so$ will be of class $C^\un$, if
$\tfrac{dt}{dr}$ can be viewed as a smooth and even function in $r=-e^f$.

\cvm
We shall assume that the function $\mu=\mu(r)$ in \re{0.18} is either even and
smooth in $(-r_0,r_0)$ 
\begin{equation}\lae{2.10}
\mu\in C^\un((-r_0,r_0))\q\tup{is even}
\end{equation}
or that $\mu\in C^\un((-r_0,0])$ such that
\begin{equation}\lae{2.11}
\abs{D^m\m(0)}=0\q\A\; m\in\N.
\end{equation}
In the latter case we can extend $\mu$ as an even and smooth function to $r>0$ by
setting
\begin{equation}
\mu(r)=\mu(-r)\q\tup{for}\q r>0.
\end{equation}

Then we can prove

\bt
Let $N$ be a brane of class $(B)$ satisfying an equation of state with
$\om=\om_0+\lam$, so that $\lam$ vanishes at the singularity. The corresponding function
$\mu$ should satisfy the conditions \re{2.10} or \re{2.11}. Then
$N$ can be reflected as described above to yield a brane $\hat N\su
\mc{\hat N}$. The joint branes
$N\uu
\hat N$ form a $C^\un$- hypersurface in $\R[2]\ti\so$ provided the following
conditions are valid

\cvm
\tup{(i)} If $\tilde\ga=\tfrac{n-1}2$ and $\om_0\le -\tfrac{n-1}2$, then the
relations
\begin{equation}\lae{2.13}
\tfrac{n-1}2\; odd,\q \om_0\; odd,\q \tup{and}\q \s\in\R[],
\end{equation}
or
\begin{equation}\lae{2.14}
n\; odd,\q \om_0\in\Z,\q \tup{and}\q \s=0
\end{equation}
should hold.

\cvm
\tup{(ii)} If $\tilde\ga=n+\om_0-1$ and $\om_0>-\tfrac{n-1}2$, then the relations
\begin{equation}
n\; odd,\q \om_0\; odd,\q \tup{and}\q \s\in\R[],
\end{equation}
or
\begin{equation}
n\; odd,\q \om_0\in\Z.\q \tup{and}\q \s=0
\end{equation}
should be valid.
\et

\bp
\cq{(i)}\q Let $\tilde\ga=\tfrac{n-1}2$ and $\om_0\le -\tfrac{n-1}2$. We have to show
that $\tfrac{dt}{dr}$ is a smooth and even function in $r$. From equation
\re{2.6} we deduce
\begin{equation}
\tfrac{dt}{dr}=-\tfrac\ka{n}(\s(-r)^{(n+\tilde\ga)f}+\rho_0e^{(\tilde\ga
-\om_0)f}e^{-\mu})\vt^{-1}\f^{-1},
\end{equation}
where
\begin{equation}\lae{2.18}
\begin{aligned}
\f^2=\abs{f'}^2e^{2\tilde\ga f}&=m e^{(2\tilde\ga  -(n-1))f}+\tfrac2{n(n+1)}\Lam
e^{2(\tilde\ga+1)f}-\tilde\ka e^{2\tilde\ga f}\\[\cma]
&\hp{=}+\tfrac{\ka^2}{n^2}(\rho_0^2e^{2(\tilde\ga+1
-(n+\om_0))f}e^{-2\mu}\\[\cma] &\qq
+
2\s\rho_0e^{(2\tilde\ga+2-(n+\om_0))f}e^{-\mu}+\s^2e^{2(\tilde\ga +1)f}).
\end{aligned}
\end{equation}

The right-hand side of \re{2.18} has to be an even and smooth function in $r=-e^f$.
By assumption $\mu$ is already an even and smooth function or can be extended to
such a  function, hence we have to guarantee that the exponents of $e^f$ are all
even.

\cvm
A thorough examination of the exponents reveals that this is the case, if the
conditions \re{2.13} or \re{2.14} are fulfilled. Notice that
\begin{equation}
\tfrac{n-1}2\; odd \im n\; odd \q\wedge \q\tfrac{3n-1}2\; even.
\end{equation}

\cvm
\cq{(ii)}\q Identical proof.
\ep

\section{Existence of the branes}

We want to prove that for the specified values of  $\om_0$ and corresponding
$\tilde\ga$ embedded branes $N$ in the black hole region of $\mc N$ satisfying the Israel
junction condition do exist.

Evidently it will be sufficient to solve the modified Friedmann equation \re{1.16} on an
interval $I=(-a,0]$ such that, if we set $r=-e^f$, then
\begin{equation}\lae{3.1}
\lim_{\tau\ra 0}r(\tau)=0,\q \lim_{\tau\ra-a}r(\tau)=r_0,
\end{equation}
where $r_0$ is the (negative) black hole radius, i.e., $\{r=r_0\}$ equals the horizon.

\cvm
We shall assume the most general equation of state that we considered in the previous
sections, i.e., $\om$ should be of the form $\om=\om_0+\lam(f)$, \cf \rl{0.2}, and the
corresponding primitive $\mu=\mu(r)$ should be smooth in $r_0<r<0$.

\cvm
Let us look at equation \re{1.18}. Setting $\f=e^{\tilde\ga f}$ we deduce that it can be
rewritten as
\begin{equation}
\tilde\ga^{-2}\dot\f^2=F(\f),
\end{equation}
where $F=F(t)$ is defined on an interval $J=[0\le t<t_0)$ and $t_0$ is given by
\begin{equation}
t_0=(-r_0)^{\tilde\ga}.
\end{equation}

$F$ is continous in $J$ and smooth in $\inn J$. Furthermore, it satisfies
\begin{equation}
F>0\q\tup{and}\q F(0)=\tilde m>0.
\end{equation}

\cvm
If we succeed in solving the equation
\begin{equation}\lae{3.5}
\tilde\ga^{-1}\dot\f=-\sqrt {F(\f)}
\end{equation}
with initial value $\f(0)=0$ on an interval $I=(-a,0]$ such that $\f\in C^0(I)\ii C^\un(\inn
I)$ and such that  the relations \re{3.1} are valid for $r=-\f^{\tilde\ga^{-1}}$, then the
modified Friedmann equation is solved by $f=-\tilde\ga^{-1}\log\f$ and $f$ satisfies
\begin{equation}\lae{3.6}
\lim_{\tau\ra 0}f=-\un,\q -f'>0.
\end{equation}

\cvm
To solve \re{3.5} let us make a variable transformation $\tau\ra-\tau$, so that we have to
solve
\begin{equation}\lae{3.7}
\tilde\ga^{-1}\dot\f=\sqrt{F(\f)}
\end{equation}
with initial value $\f(0)=0$ on an interval $I=[0,a)$.

\bt\lat{3.1}
The equation \re{3.7} with initial value $\f(0)=0$ has a solution $\f\in C^1(I)\ii C^\un(\inn
I)$, where $I=[0,a)$ is an interval such that the relations \re{3.1} are satisfied by
$r=-\f^{\tilde\ga^{-1}}$ with obvious modifications resulting from the variable
transformation $\tau\ra-\tau$.
\et

\bp
Let $0<t'<t_0$ be arbitrary and define $J'=[0,t']$. Then there are positive constants
$c_1,c_2$ such that
\begin{equation}\lae{3.8}
c_1^2\le F(t)\le c_2^2\qq\A\; t\in J'.
\end{equation}

Let $\e>0$ be small, then the differential equation \re{3.7} with initial value $\f_\e(0)=\e$
has a smooth, positive solution $\f_\e$ in $0\le \tau< \tau_\e$, where $\tau_\e$ is
determined by the requirement
\begin{equation}
\f_\e(\tau)\le t'\qq\A\; 0\le\tau< \tau_\e.
\end{equation}

In view of the estimates \re{3.8} we immediately conclude that
\begin{equation}
\tilde\ga c_1\tau\le\f_\e(\tau)\le\e+ \tilde\ga c_2\tau\qq\A\; 0\le\tau<\tau_\e,
\end{equation}
hence, if we choose the maximal $\tau_\e$ possible, then there exists $\tau_0>0$ such
that
\begin{equation}
\tau_\e\ge \tau_0>0\qq\A\; \e.
\end{equation}

\cvm
Now, using well-known a priori estimates, we deduce that for any interval $I'\Su (0,\tau_0)$
we have
\begin{equation}
\abs{\f_\e}_{m,I'}\le c_m(I')\qq\A\; m\in\N
\end{equation}
independently of $\e$. Moreover, since $\f_\e$ is uniformly Lipschitz continous in
$[0,\tau_0]$, we infer that
\begin{equation}
\lim_{\e\ra0}\f_\e=\f
\end{equation}
exists such that $\f\in C^1([0,\tau_0])\ii C^\un((0,\tau_0))$ and $\f$ solves \re{3.7}
with initial value $\f(0)=0$.

\cvm
$\f$ can be defined on a maximal interval interval $[0,a^*)$, where $a^*$ is determined by
\begin{equation}
\lim_{\tau\ra a^*}\f(\tau)=\un\q\tup{or}\q \lim_{\tau\ra a^*}F(\f)=0.
\end{equation}

Obviously, there exists $0<a\le a^*$ such that
\begin{equation}
\lim_{\tau\ra a}\f(\tau)=(-r_0)^{\tilde\ga}.
\end{equation}
Hence the proof of the theorem is completed.
\ep

\br
If $a<a^*$, then we cannot conclude that we have defined branes extending beyond the
horizon, since our embedding deteriorates when the horizon is approached, as a look at the
equations \re{1.15} and \re{1.16} reveals. Since $\tilde h$ vanishes on the horizon, either
$\tfrac{dt}{d\tau}$ becomes unbounded or $(\rho+\s)$ would tend to zero, in which case
$F(\f)$ would vanish, i.e., if $a<a^*$, then necessarily
\begin{equation}
\lim_{\tau\ra-a}\abs{\tfrac{dt}{d\tau}}=\un
\end{equation}
has to be valid. 
 An extension of the brane past the horizon will require a different embedding or at least a
different coordinate system in the bulk. Notice that
\begin{equation}
\bar g_{00}=\spd {\dot y}{\dot y}=-r^2\ne 0,
\end{equation}
so that it is more or less a coordinate singularity.
\er

\nocite{khoury:cyclic,steinhardt:cyclic,cg:imcf,langlois:brane}
\bibliographystyle{amsplain}

\begin{thebibliography}{1}

\bibitem{cg:indiana}
Claus Gerhardt, \emph{Hypersurfaces of prescribed curvature in {L}orentzian
  manifolds}, Indiana Univ. Math. J. \textbf{49} (2000), 1125\ndash 1153,
  \href{http://www.math.uni-heidelberg.de/studinfo/gerhardt/bibtexcg00b.html}{%
pdf file}.

\bibitem{cg:arw}
\bysame, \emph{The inverse mean curvature flow in {ARW} spaces - transition
  from big crunch to big bang}, 2004, 39 pages,
  \href{http://arXiv.org/pdf/math.DG/0403485}{math.DG/0403485}.

\bibitem{cg:imcf}
\bysame, \emph{The inverse mean curvature flow in cosmological spacetimes},
  2004, 24 pages, \href{http://arxiv.org/pdf/math.DG/0403097}{math.DG/0403097}.

\bibitem{cg:rw}
\bysame, \emph{The inverse mean curvature flow in {Robertson-Walker} spaces and
  its application to cosmology}, 2004, 9 pages,
  \href{http://arXiv.org/pdf/gr-qc/0404112}{gr-qc/0404112}.

\bibitem{cg:mass}
\bysame, \emph{The mass of a {L}orentzian manifold}, 2004, 14 pages,
  \href{http://arXiv.org/pdf/math.DG/0403002}{math.DG/0403002}.

\bibitem{khoury:cyclic}
Justin Khoury, \emph{A briefing on the ekpyrotic/cyclic universe}, 2004, 8
  pages, \href{http://arXiv.org/pdf/astro-ph/0401579 }{astro-ph/0401579}.

\bibitem{langlois:brane}
David Langlois, \emph{Brane cosmology: an introduction}, 2002, 32 pages, \href{
  http://arXiv.org/pdf/hep-th/0209261}{hep-th/0209261}.

\bibitem{steinhardt:cyclic}
Neil Turok and Paul~J. Steinhardt, \emph{Beyond inflation: A cyclic universe
  scenario}, 2004, 27 pages,
  \href{http://arXiv.org/pdf/hep-th/0403020}{hep-th/0403020}.

\end{thebibliography}
\providecommand{\bysame}{\leavevmode\hbox to3em{\hrulefill}\thinspace}
\providecommand{\MR}{\relax\ifhmode\unskip\space\fi MR }
\providecommand{\MRhref}[2]{%
  \href{http://www.ams.org/mathscinet-getitem?mr=#1}{#2}
}
\providecommand{\href}[2]{#2}



\end{document}